\documentstyle[12pt]{article} 
\begin{document}
\renewcommand{\theequation}{\thesection .\arabic{equation}}
\newcommand{\beq}{\begin{equation}}
\newcommand{\eeq}{\end{equation}}
\newcommand{\beqn}{\begin{eqnarray}}
\newcommand{\eeqn}{\end{eqnarray}}
\newcommand{\slp}{\raise.15ex\hbox{$/$}\kern-.57em\hbox{$\partial
$}}
\newcommand{\lnA}{\raise.15ex\hbox{$/$}\kern-.57em\hbox{$A$}}
\newcommand{\lnB}{\raise.15ex\hbox{$/$}\kern-.57em\hbox{$B$}}
\newcommand{\bP}{\bar{\Psi}}
\newcommand{\slU}{\raise.15ex\hbox{$/$}\kern-.57em\hbox{$U$}}
\newcommand{\hfi}{\hat{\Phi}} 
\newcommand{\heta}{\hat{\eta}}
\newcommand{\hv}{\hat{V^{ab}}}
\newcommand{\hs}{\hspace*{0.6cm}}

\title{Equivalent bosonic theory for the massive Thirring
model with non-local interaction}

\author{Kang Li$^{a,b}$ and Carlos Na\'on$^{a}$}
\date{June 1997}
\maketitle
\begin{abstract}
We study, through path-integral methods, an extension of the
massive Thirring model in which the interaction between currents is non-local.
By examining the mass-expansion of the partition function we show that this
non-local massive Thirring model is equivalent to a certain non-local extension
of the sine-Gordon theory. Thus, we establish a non-local generalization of the
famous Coleman's equivalence. We also discuss some possible applications of 
this result in the context of one-dimensional strongly correlated systems and
finite-size Quantum Field Theories. 
\end{abstract}

\vspace{3cm}
Pacs: 11.10.Lm\\ 
\hspace*{1,7 cm} 05.30.Fk

\noindent --------------------------------

\noindent $^a$ {\footnotesize Depto. de F\'\i sica.  Universidad
Nacional de La Plata.  CC 67, 1900 La Plata, Argentina.\\
E-Mail: naon@venus.fisica.unlp.edu.ar}

\noindent $^b$ {\footnotesize Department of Physics, Hangzhou
University, Hangzhou, 310028, P. R. China.}

\newpage

\section{Introduction}

\hs Bosonization, i.e. the equivalence between fermionic and bosonic Green
functions in $1+1$ dimensions, has by now a long history \cite{Stone}, that 
could be traced back to the work of F.Bloch on the energy loss of charged 
particles travelling through a metal \cite{Bloch}. In more recent years, 
the famous Coleman's equivalence proof between the massive Thirring and
Sine-Gordon theories \cite{Coleman} and Polyakov and Wiegman and Witten's 
non-Abelian bosonization \cite{Witten}, helped to convert this procedure into 
a standard and powerful tool for the understanding of Quantum Field Theories 
(QFT's). All these achievements were realized in the context of {\em local} 
QFT's. 

\hs Recently, the bosonization procedure in its path-integral version 
\cite{N} was applied, for the first time, to a {\em non-local} QFT, 
namely, a Thirring model with massless fermions and a non-local 
(and non-covariant) interaction between fermionic currents \cite{NR}.
The study of such a model is relevant, not only from a purely field-theoretical
point of view but also because of its connection with the physics of strongly
correlated systems in one spatial dimension (1d). Indeed, this model describes
an ensemble of non-relativistic particles coupled through a 2-body forward-
scattering potential and displays the so called Luttinger-liquid behaviour
\cite{Hal} that could play a role in real 1d semiconductors 
(see for instance \cite{Tans}).

\hs In this paper we undertake the path-integral bosonization of the non-local
Thirring model (NLT) with a relativistic fermion mass term included in the 
action. Using a functional decoupling technique to treat the non-locality
\cite{NR}, and performing a perturbative expansion in the mass parameter,
we find  that the NLT is equivalent to a purely bosonic action which is a
simple non-local extension of the sine-Gordon model. Thus, our main result can
be considered as a generalization of Coleman's equivalence to the case in which
the usual Thirring interaction is point-splitted through bilocal potentials.
In the language of many-body, non-relativistic systems, the relativistic mass
term can be shown to represent not an actual mass, but the introduction of
backward-scattering effects \cite{Back}. Therefore, our result provides an 
alternative route to explore the dynamics of collective modes in 1d strongly
correlated systems.

\hs The paper is organized as follows. In Section 2 we present the model and
write the partition function in terms of a massive fermionic determinant. Then
we perform a perturbative expansion in the mass parameter and evaluate every
free (massless) vacuum expectation value (v.e.v.). This allows us to obtain
an explicit expression for the partition function of the massive NLT.\\
In Section 3 we introduce a modified sine-Gordon model (NLSG) with an additional
non-local term. We then consider the corresponding partition 
function, make an expansion in the cosine term and compute the free v.e.v.'s,
for each term in the series. Finally we compare both the fermionic and 
bosonic expansions term by term and find that they are equal if a certain
relationship between NLT and NLSG potentials is satisfied. We end this Section
by briefly showing how our approach can be used to study the 1d electronic 
liquid with back-scattering. We also comment on the possibility of exploiting
this work in order to shed some light on the validity of Coleman's equivalence
at finite volume.\\
In the last Section we summarize and stress the main aspects of this work.

\section{Partition function for the massive non-local Thirring model}
\setcounter{equation}{0}

\hs Let us consider the Lagrangian density of the massive Thirring model with 
a non-local interaction between fermionic currents

\beq
L=i\bar{\Psi}\slp\Psi+\frac{1}{2}g^2\int d^2 yJ_{\mu}(x)V_{(\mu )}(x,y)
J_{\mu}(y) -m\bP\Psi ,
\label{2}
\eeq

\noindent where
\beq
J_{\mu}(x)=\bP (x)\gamma_{\mu}\Psi (x) ,
\label{3}
\eeq
\noindent and
\beq     
V_{(\mu)}(x,y)=V_{(\mu)}(|x-y|)
\label{4}
\eeq
\noindent is an arbitrary function of $|x-y|$.  
 We shall use $\gamma$ matrices defined as
\beq
\gamma_0=\left(\begin{array}{cc}
0&1\\
1&0
\end{array}\right), \gamma_1=\left(\begin{array}{cc}
0&i\\
-i&0
\end{array}\right),
\gamma_5=i\gamma_0\gamma_1=\left(\begin{array}{cc}
1&0\\
0&-1
\end{array}\right)
\eeq

\beq
[\gamma_{\mu},\gamma_{\nu}]=2\delta_{\mu\nu},
~~~~ \gamma_{\mu}\gamma_5=i\epsilon_{\mu\nu}\gamma_{\nu}.
\eeq

The partition function
of the model is 

\beq
Z=N\int D\bP D\Psi e^{-\int d^2x L}
\label{5}
\eeq
By using the following representation of the functional delta,

\beq
\delta (C_{\mu})=\int D\tilde{A_{\mu}}exp(-\int d^2x\tilde{A_{\mu}}
C_{\mu}),
\eeq

\noindent we can write $Z$ as 

\beq
Z = N \int D\bP D\Psi D\tilde{A}_{\mu}D\tilde{B}_{\mu}~ exp\{-\int
d^2x [\bP (i\slp -m)\Psi +\tilde{A}_{\mu}\tilde{B}_{\mu}
+ \frac{g}{\sqrt{2}}(\tilde{A}_{\mu}
J_{\mu} + \tilde{B}_{\mu}K_{\mu})] \}
\label{8}
\eeq      

\noindent where
\beq
K_{\mu}(x)=\int d^2 y V_{(\mu)}(x,y)J_{\mu}(y).
\eeq

\noindent Please note that no sum over repeated indices
is implied when a subindex $(\mu)$ is involved.

If we define
\beq
\bar{B}_{\mu}(x) = \int d^2y~ V_{(\mu)}(y,x)\tilde{B}_{\mu}(y),
\label{11}
\eeq
\beq
\tilde{B}_{\mu}(x) = \int d^2y~ b_{(\mu)}(y,x) \bar{B}_{\mu}(y),
\label{12}
\eeq

\noindent with $b_{(\mu)}(y,x)$ satisfying  

\beq
\int d^2y~ b_{(\mu)}(y,x) V_{(\mu)}(z,y) = \delta^2 (x-z),
\label{13}
\eeq

\noindent and change auxiliary variables in the form

\beq
A_{\mu}=\frac{1}{\sqrt{2}}(\tilde{A}_{\mu} +\bar{B}_{\mu}),
\label{15a}
\eeq       
    
\beq
B_{\mu}=\frac{1}{\sqrt{2}}(\tilde{A}_{\mu} - \bar{B}_{\mu}),
\label{15}
\eeq

\noindent we obtain
\beq
Z = N \int DA_{\mu} DB_{\mu}~ det(i \slp + g \lnA -m) e^{-S(A,B)},
\label{16}
\eeq

\noindent where 
\beqn
S(A,B)=\frac{1}{2}\int d^2x\int d^2y b_{(\mu)}(x,y)[A_{\mu}(x)
      A_{\mu}(y)-B_{\mu}(x)B_{\mu}(y)]
\label{17}
\eeqn

\noindent The Jacobian associated with the change $(\tilde{A}, \tilde
{B})\rightarrow (A,B)$ is field- independent and can then be absorbed in
the normalization constant $N$.
 From (\ref{16}) and (\ref{17}) we find that the fields
$B_{\mu}$ are completely decoupled from both fermion fields and $A_{\mu}$
fields, so their contribution can be also factorized and 
absorbed in $N$. Thus we can now write $Z$ as

\beq
Z = N \int DA_{\mu} det(i \slp + g \lnA -m) e^{-S[A]},
\label{18}
\eeq

\noindent with
\beq
S(A)=\frac{1}{2}\int d^2xd^2y b_{(\mu)}(x,y)A_{\mu}(x)A_{\mu}(y).
\eeq

As it is known, the massive determinant in (\ref{18}) has not been
exactly solved yet. The usual way of dealing with it consists in performing
a chiral transformation in the fermionic path-integral variables and making
then an expansion with $m$ as perturbative parameter (This 
procedure was employed for the local case in ref.\cite{N}). 
Since we were able to write the partition function in such a way that non-local
terms are not present in the fermionic determinant, we can follow exactly the 
same strategy as in the local case. To this aim let us first express the vector 
field in terms of two new fields $\Phi$ and $\eta$ as

\beq
A_{\mu}(x)=-\epsilon_{\mu\nu}\partial_{\nu}\Phi (x)+
\partial_{\mu}\eta (x),
\eeq

\noindent which can be considered as a change of bosonic variables with
trivial (field-independent) Jacobian. We also make the change

\beq
\Psi(x) = exp[-g(\gamma_5 \Phi(x) + i \eta(x))] \chi(x)
\label{19}
\eeq

\beq
\bP(x) = \bar\chi(x) exp[-g(\gamma_5 \Phi(x) - i\eta(x))]
\label{19a}
\eeq

\noindent with non-trivial Jacobian given by 
\beq
J_F =exp[\frac{g^2}{2\pi}\int d^2x\Phi (x)\Box \Phi (x)]
\eeq

\noindent Then we get

\beq
Z=N\int D\bar\chi D\chi D\Phi D\eta e^{-S_{eff}}
\label{23}
\eeq
\noindent where
\beq
S_{eff}=S_{0F}+S_{0NLB}-m\int d^2x\bar\chi e^{-2g\gamma_5\Phi}\chi,
\eeq

\beq
S_{0F}=\int d^2x(\bar\chi i\slp\chi ),
\eeq
\noindent and
\beqn
 S_{0NLB}=\frac{g^2}{2\pi} 
           \int d^2x~ (\partial_{\mu}\Phi)^2 \nonumber\\
        &+ \frac
        {1}{2}\int d^2x d^2y \epsilon_{\mu\lambda}\epsilon_{\mu\sigma}
        b_{(\mu)}(y,x) \partial_{\lambda} \Phi (x)
        \partial_{\sigma} \Phi (y) \nonumber\\
        &+\frac{1}{2}\int d^2x d^2y b_{(\mu)}(y,x) \partial_{\mu}
        \eta (x)
        \partial_{\mu}\eta (y) \nonumber\\
        &-\int d^2x d^2y [b_{(0)}(y,x) \partial_0 \eta (x)
        \partial_1 \Phi (y) \nonumber\\
        &- b_{(1)}(y,x) \partial_1 \eta (x)
        \partial_0 \Phi (y)]
\label{28}
\eeqn

\noindent Note that for
\beq
\partial_1^x\partial_0^y b_0(x,y)=\partial_0^x\partial_1^y b_1(x,y)\\
\eeq

\noindent the last term of $S_{0NLB}$ vanishes, and in this case
$\eta (x)$ decouples from $\bar\chi,~\chi$ and $\Phi$.
In the general case $S_{0NLB}$ describes a system of two bosonic fields coupled 
by distance-dependent coefficients.

Exactly as one does in the local case, the partition function for the massive
NLT can be formally written as a mass-expansion:

\beq
Z=\sum_{n=0}^{\infty}\frac{(m)^n}{n!}<\prod_{j=1}^n \int d^2 x_j 
\bar\chi (x_j) e^{-2g\gamma_5\Phi (x_j)}\chi (x_j )>_{0}
\label{33}
\eeq

\noindent where the $<~>_{0}$ means v.e.v.
in the theory of free massless fermions and non-local bosons.
Using now the identity

\beq
\bar\chi (x_j) e^{-2g\gamma_5\Phi (x_j)}\chi (x_j )
=e^{-2g\Phi}\bar\chi\frac{1+\gamma_5}{2}\chi +
e^{2g\Phi}\bar\chi\frac{1-\gamma_5}{2}\chi,
\eeq

\noindent equation (\ref{33}) can be written as

\beqn
Z=&\sum_{k=0}^{\infty}\frac{(m)^{2k}}{{k!}^2}\int\prod_{i=1}^k d^2 x_i
d^2y_i <exp[2g\sum_i (\Phi (x_i)-\Phi (y_i))]>_{0NLB}\nonumber\\
~&\cdot <\prod_{i=1}^k \bar\chi (x_i)\frac{1+\gamma_5}{2}\chi (x_i)
\bar\chi (y_i)\frac{1-\gamma_5}{2}\chi (y_i)>_{0F}
\label{34}
\eeqn

\noindent Each fermionic part can be readily computed by writing

\beqn
\bar\chi\frac{1+\gamma_5}{2}\chi =\bar{\chi_1}\chi_1 \nonumber\\
\bar\chi\frac{1-\gamma_5}{2}\chi =\bar{\chi_2}\chi_2,
\eeqn

\noindent where $ \chi=(\begin{array}{c}\chi_1 \\
\chi_2\end{array} ), \bar\chi =(\bar{\chi_1} , \bar{\chi_2})$, and using
Wick's theorem with the usual free fermion propagator.

Concerning the bosonic (non-local) factors, they are more easily handled in
momentum space. The corresponding Fourier transformed action acquires the
following more compact form:

 \beqn
 S_{0NLB}& =& \frac{1}{(2\pi)^2} \int d^2p \{\hat{\Phi}(p) \hat{\Phi}(-p)
           A(p) \nonumber \\
           &+& \hat{\eta}(p) \hat{\eta}(-p) B(p) -\hat{\Phi}(p)
           \hat{\eta}(-p) C(p) \},
\label{29}
\eeqn

\noindent where

 \beq
     A(p) = \frac{g^2}{2\pi}~ p^2 + 
     \frac{1}{2}[\hat{b}_{(0)}(p) p_1^2 +
           \hat{b}_{(1)}(p) p_0^2],
\label{30}
\eeq

\beq
B(p) = \frac{1}{2}[\hat{b}_{(0)}(p) p_0^2 +
           \hat{b}_{(1)}(p) p_1^2],
\label{31}
\eeq

\beq
C(p) = [\hat{b}_{(0)}(p) - \hat{b}_{(1)}(p)] p_0 p_1,
\label{32}
\eeq
                                         
\noindent $p^2 = p_0^2 + p_1^2$,
and $\hat{\Phi},\hat{\eta}$ and $\hat{b}_{(\mu)}$ are the Fourier
transforms of $\Phi, \eta$ and $b_{(\mu)}$ respectively.

One then has
\beqn
 <exp[2g\sum_i (\Phi (x_i)-\Phi (y_i))]>_{0NLB}
= \nonumber\\
\frac{\int D\hat{\Phi}(p)D\heta (p)e^{-S_{0NLB}}\cdot 
e^{\frac{g}{\pi}\sum_i\int d^2p \Phi (p) (e^{ipx_i}-e^{ipy_i})}}
{\int D\hat{\Phi}(p)D\heta (p)e^{-S_{0NLB}}}
\eeqn

\noindent This v.e.v. can be computed by translating the quantum fields 
$\hat{\Phi}(p)$ and $\heta (p)$, 

\beqn
\hat{\Phi}(p)&=\hat{\phi}(p) +E(p) \nonumber\\
\heta (p)& =\hat{\rho}(p) +F(p)
\eeqn

\noindent where $\hat{\phi}$ and $\hat{\rho}$ are the new quantum fields,
whereas $E(p)$ and $F(p)$ are two classical functions satisfying

\beqn
E(-p)=-\frac{4 g B(p)}{\Delta (p)}D(p,x_i,y_i) \nonumber \\
F(-p)=-\frac{2 g C(p)}{\Delta (p)}D(p,x_i,y_i) 
\eeqn

\noindent with

\beq
\Delta = C^2(p)-4A(p)B(p)
\eeq
and
\beq
D(p,x_i,y_i) =\sum_i (e^{ipx_i}-e^{ipy_i})
\label{37}
\eeq
           
\noindent We then get

\beqn
 <exp[2g\sum_i (\Phi (x_i)-\Phi (y_i))]>_{0NLB}
=\nonumber\\
exp\{-(\frac{g}{\pi})^2\int d^2p\frac{B(p)}{\Delta (p)}D(p,x_i,y_i)D(-p,x_i,y_i) \}
\eeqn

Gathering this result together with the Fourier transformed fermionic factors
\cite{B}, we find
\beqn
Z&=\sum_{k=0}^{\infty}\frac{(m)^{2k}}{(k!)^2}\int\prod_{i=1}^k d^2x_i 
d^2y_i exp\{ -\int \frac{d^2p}{(2\pi)^2}[\frac{2\pi}{p^2}\nonumber\\
&-\frac{2\pi g^2(\hat{b}_{(0)}p_0^2+\hat{b}_{(1)}p_1^2 )}
{ g^2(\hat{b}_{(0)}p_0^2+\hat{b}_{(1)}p_1^2)p^2+\pi\hat{b}_{(0)}
\hat{b}_{(1)}p^4}]D(p,x_i,y_i)D(-p,x_i,y_i)\}
\label{49}
\eeqn

Thus we have been able to obtain an explicit expansion for the partition
function of a massive Thirring model with arbitrary (symmetric) bilocal 
potentials coupling the fermionic currents. This result will be used in
the next Section in order to establish, by comparison, its equivalence to
a sine-Gordon-like model.

\section{Connection with a non-local sine-Gordon model}
\setcounter{equation}{0}

\hs Let us now consider the Lagrangian density of the non-local sine-Gordon 
model (NLSG) given by 

\beq
{\cal L}_{NLSG} =\frac{1}{2}(\partial_{\mu}\phi(x))^2+\frac{1}{2}
\int d^2y\partial_{\mu}\phi(x)d_{(\mu )}(x-y)\partial_{\mu}\phi(y)
-\frac{\alpha_0}{\beta^2}\cos \beta\phi
\label{3.1}
\eeq

\noindent where $d_{(\mu )}(x-y)$ is an arbitrary potential function
of $|x-y|$. The partition function of this model reads 

\beq
Z_{NLSG}=\int D\phi exp[-\int d^2x {\cal L}_{NLSG} ].
\label{3.2}
\eeq

\noindent Performing a perturbative expansion in $\alpha_0$, we obtain

\beq
Z_{NLSG}=\sum_{k=0}^{\infty}[\frac{1}{k!}]^2 (\frac{\alpha_0}
{\beta^2})^{2k}\int\prod_{i=1}^k d^2x_i d^2y_i <e^{i\beta\sum_i 
(\phi(x_i)-\phi(y_i))} >_{0},
\label{3.3}
\eeq 

\noindent where $<~~>_{0}$ means the v.e.v. with respect to the "free" action
defined by the two first terms in the right hand side of equation (\ref{3.1}). 
Since we are again led to the computation of v.e.v.'s of vertex operators, from
now on the technical aspects of the calculation are, of course, very similar 
to the ones depicted in the previous Section. For this reason we shall omit the
details here. The result is
\newpage
\beqn 
Z_{NLSG}=\sum_{k=0}^{\infty}[\frac{1}{k!}]^2 (\frac{\alpha_0}
{\beta^2})^{2k}\int\prod_{i=1}^k d^2x_i d^2y_i \nonumber\\
.~ exp[-\frac{\beta^2}{4}
\int \frac{d^2p}{(2\pi)^2}\frac{D(p,x_i,y_i)D(-p,x_i,y_i)}
              {\frac{1}{2}p^2+\frac{1}{2}(\hat{d}_{(0)}(p) p_0^2+
\hat{d}_{(1)}(p) p_1^2)  }]
\label{3.4}
\eeqn

By comparing equation (\ref{49} ) with equation (\ref{3.4}), 
we find that both expansions are identical if the following equations hold:
\beq 
m=\frac{\alpha_0}{\beta^2}
\label{56}
\eeq
and

\beq 
\frac{1}
{\frac{g^2}{\pi}(\frac{p_0^2}{\hat{b}_{(1)}}
+\frac{p_1^2}{\hat{b}_{(0)}}) + p^2}=
\frac{\beta^2}
{4\pi(p^2+ \hat{d}_{(0)} p_0^2+
\hat{d}_{(1)} p_1^2) }
\label{3.5}
\eeq 
where, for the sake of clarity, we have omitted the p-dependence of the
potentials.

Therefore, we have obtained a formal equivalence between the partition 
functions of the massive NLT and NLSG models. This is the main result of this
paper.

In order to check the validity of equation (\ref{3.5}), let us specialize 
it to the covariant case,
\beqn 
\hat{b}_{(0)}(p)=\hat{b}_{(1)}(p)=\hat{b}(p)\nonumber\\
\hat{d}_{(0)}(p)=\hat{d}_{(1)}(p)=\hat{d}(p)
\label{57}
\eeqn

\noindent which yields

\beq 
\frac{1}{\frac{g^2}{\pi \hat{b}(p)}+ 1}
=\frac{\beta^2}{4\pi (1+\hat{d}(p))}
\label{3.8}
\eeq

In particular, when $\hat{b}(p) =1$ and $\hat{d}(p)=0$, our massive NLT
model returns to the usual massive Thirring model, and the
NLSG model becomes the ordinary sine-Gordon model. Making these replacements
in equation (\ref{3.8}) we get

\beq 
\frac{\beta^2}{4\pi}=\frac{1}{1+\frac{g^2}{\pi}}
\label{59}
\eeq

\noindent which is the well-known Coleman's result (\cite{Coleman}).
Of course, in this particular case one also has a modified version of the
identity (\ref{56}), with both $m$ and $\alpha_{0}$ renormalized due to
divergencies coming from the vertex operators v.e.v.'s.

It is certainly encouraging to reproduce equation (\ref{59}). However, our more
general formula (\ref{3.5}) enables to profit from the bosonization 
identification in a much wider variety of situations, and in a very 
straightforward way. In particular, the non-local version of the sine-Gordon 
model can be easily used in the context of 1d strongly correlated fermions. This 
type of systems has attracted a lot of attention in the last years, due to 
striking advances in the material sciences that have allowed to build real 
"quantum wires" \cite{QW}. Much of the theoretical understanding of these 
physical systems has come from the study of the Tomonaga-Luttinger (TL) model 
\cite{ML} \cite{S} \cite{Voit} which, in its simpler version, describes spinless 
fermions interacting through their density fluctuations. In ref.\cite{NR} it
has been shown that the TL model is a particular case of the NLT model
considered in the present work, corresponding to 
$\hat{b}_{(1)}\rightarrow\infty$ and $\hat{b}_{(0)}$ associated to the density-
density interaction, $V(p_1) = \frac{1}{\hat{b}_{(0)}(p)}$. For this many-body
system, adding a relativistic fermion mass is intimately connected to the
description of backward-scattering processes (the so-called Luther-Emery model
\cite{Back}). Therefore, the NLSG model could be used to explore the 
Luther-Emery model. For illustrative purposes we shall consider here the 
spinless case, although the extension to the spin-$\frac{1}{2}$ case can be 
easily done within this framework (See \cite{NR}).
To do this, according to the previous discussion, one has to take the limit
$\hat{b}_{(1)}\rightarrow\infty$ in (\ref{3.5}), thus obtaining
\beq 
\frac{1}
{\frac{g^2}{\pi}V(p_1)p_1^2 + p^2}=
\frac{\beta^2}
{4\pi(p^2+ \hat{d}_{(0)} p_0^2+
\hat{d}_{(1)} p_1^2) }
\eeq 
In this context the above equation has to be viewed as an identity that
permits to determine the potentials $\hat{d}_{\mu}$ necessary to analyze the
original fermionic Luther-Emery model in terms of the bosonic NLSG model. For 
instance, if we set $\hat{d}_{(0)} = 0$ and $\beta^2 = 4\pi$, $\hat{d}_{(1)}$ 
turns out to be proportional to $V$. Then equation (\ref{3.1}) describes the 
dynamics of the collective modes, whose spectrum, as it is well-known, 
develops a gap due to back-scattering effects. By virtue of equation (\ref{56})
one can directly read the value of this gap from (\ref{3.1}), obtaining
$m = \frac{\alpha_0}{4\pi}$.\\
In passing let us mention that the study of the Luther-Emery model in the 
presence of impurities could be also undertaken by combining the present scheme 
with the results of ref.\cite{Imp}.\\
We think that the identification we established in this paper might be useful 
to compute finite size corrections in the massive Thirring model. There has 
been some recent interesting studies on this subject which found different 
results for the values of the central charges of the massive Thirring \cite{FT}
and sine-Gordon \cite{EY} models. One possible explanation for this 
disagreement has been given in ref.\cite{KM}, where it is argued, by using
perturbed conformal field theory, that Coleman's equivalence is spoiled by 
finite volume effects. All these investigations are restricted to local
models. In this context, our approach could be employed to examine the 
influence of non-contact interactions on the perturbed conformal properties of
the systems. Since such a computation is expected to be closely related to
the ground-state structure of the theories under consideration, it will be 
facilitated by recent results on the vacuum properties of the NLT model 
\cite{BN}. This problem is beyond the scope of the present article, but will be
addressed in the close future. 

\newpage

\section{Summary}
\hs In this work we have considered an extension of the massive Thirring model 
in which the fermionic current-current interaction is mediated by distance 
dependent potentials. We also introduced a simple modification of the 
sine-Gordon model that consists in adding a non-local kinetic-like term to the 
usual bosonic action. By analyzing the vacuum to vacuum functionals of each
model through perturbative expansions, in complete analogy with the original
procedure followed by Coleman in his celebrated paper \cite{Coleman}, 
we found that both series (the mass expansion of the Thirring-like model and
the "fugacity" expansion of the sine-Gordon-like model) are equal provided
that a certain relation between the corresponding potentials is satisfied
This is our main result (See equation (\ref{3.5})). Taking into account the
close connection between the non-local Thirring model and a non-relativistic
many-body system of one-dimensional electrons, we have depicted how to use
our result in order to study the back-scattering
problem by means of the non-local sine-Gordon theory proposed in this paper.
We have also stressed the possibility of using our result as an alternative
tool to check the validity of Coleman's equivalence at finite volume.

\section{Acknowledgements}
One of us (K.L) would like to thank  Prof. R.Gamboa Sarav\'{\i} for
hospitality in Universidad Nacional de La Plata (UNLP), Argentina, 
where this paper is done.\\
C.N. is partially supported by Consejo Nacional de Investigaciones 
Cient\'{\i}ficas y T\'ecnicas (CONICET), Argentina.

\end{document}